\documentclass[sigplan,screen]{acmart}
\usepackage{algorithm}
\usepackage{algpseudocode}
\AtBeginDocument{%
  }

\copyrightyear{2023}
\acmYear{2023}
\setcopyright{rightsretained}
\acmConference[ICMVA 2023]{The 7th International Conference on Machine Vision and Applications (ICMVA)}{March 12--14, 2024}{Singapore}
\acmBooktitle{2024 The 7th International Conference on Machine Vision and Applications (ICMVA 2024), March 12--14, 2024, Singapore}\acmDOI{}
\acmISBN{}

\begin{document}

\title{Diffusion and Multi-Domain Adaptation Methods for Eosinophil Segmentation}

\author{Kevin Lin}
\orcid{0000-0002-4968-0532}
\affiliation{%
  \department{School of Data Science}
  \institution{University of Virginia}
  \city{Charlottesville}
  \state{Virginia}
  \country{USA}
}
\email{pex7ps@Virginia.edu}

\author{Donald Brown}
\orcid{0000-0002-9140-2632}
\affiliation{%
  \department{School of Data Science}
  \institution{University of Virginia}
  \city{Charlottesville}
  \state{Virginia}
  \country{USA}}
\email{deb@virginia.edu}

\author{Sana Syed}
\orcid{0000-0003-0954-0583}
\affiliation{%
  \department{Department of Pediatric Gastroenterology}
  \institution{University of Virginia}
  \city{Charlottesville}
  \state{Virginia}
  \country{USA}
}
\email{ss8xj@virginia.edu}

\author{Adam Greene}
\orcid{0009-0004-3979-5813}
\affiliation{%
  \department{Department of Pediatric Gastroenterology}
  \institution{University of Virginia}
  \city{Charlottesville}
  \state{Virginia}
  \country{USA}
}
\email{arg7ef@virginia.edu}

\renewcommand{\shortauthors}{Lin et al.}

\begin{abstract}
Eosinophilic Esophagitis (EoE) represents a challenging condition for medical providers today. The cause is currently unknown, the impact on a patient's daily life is significant, and it is increasing in prevalence. Traditional approaches for medical image diagnosis such as standard deep learning algorithms are limited by the relatively small amount of data and difficulty in generalization. As a response, two methods have arisen that seem to perform well: Diffusion and Multi-Domain methods with current research efforts favoring diffusion methods. For the EoE dataset, we discovered that a Multi-Domain Adversarial Network outperformed a Diffusion based method with a FID of 42.56 compared to 50.65. Future work with diffusion methods should include a comparison with Multi-Domain adaptation methods to ensure that the best performance is achieved.
\end{abstract}



\received{10 Feb 2024}
\received[revised]{XX XXXX 20XX}
\received[accepted]{XX XXXX 20XX}

\maketitle

Eosinophilic esophagitis (EoE) represents an allergic response in the esophagus creating an excess of white blood cells. This condition affects approximately 0.5-1.0 in 1,000 people and it can be seen in 2-7\% of patients that undergo endoscopies \cite{Dellon2014EpidemiologyEsophagitis}. While the exact cause of EoE is not fully understood, pathologists believe that a patient's diet can trigger this condition. In addition, the disease is increasing in prevalence \cite{Carr2018EosinophilicEsophagitis}, increasing the workload on pathologists. Furthermore, EoE directly impacts these patient's daily lives causing swallowing difficulties, food impaction, and chest pain\cite{Runge2017CausesEsophagitis}. Current treatment for EoE requires a patient to undergo an endoscopy where eosinophil biospy samples are extracted and evaluated for a concentration assessment. The current diagnostic criteria for EoE is 15 or more eosinophils in at least one High-Power Field (HPF: 400x magnification adjustment)\cite{Furuta2007EosinophilicGastroenterol}. In this study, the EoE dataset is obtained from the Gastroenterology Data Science Lab from the University of Virginia Medical Center's patient data. A sample image in the dataset of a Whole Slide Image (WSI) is given in Figure (a). The WSI is then divided into 514 smaller images/masks each with a dimension of 512x512x3. The dataset encompasses the biospy images of 30 UVA Medical Center patients, each who have been diagnosed by pathologists as having EoE. As the image has been treated with a hematoxylin and eosin stain (H\&E), the color channel of [r,g,b] will be maintained to assist further analysis while the masks will be imported in grayscale. In accordance with patient privacy laws including the United States Health Insurance Portability and Accountability Act of 1996 (HIPAA) and the University of Virginia Medical Center's own commitment to patient privacy, all data used in this study is obtained under conditions of academic use only. Additionally, no personal health information (PHI) is present in the data. 

\begin{figure}[htb]

\begin{minipage}[b]{1.0\linewidth}
  \centering
  \centerline{\includegraphics[width=8.5cm]{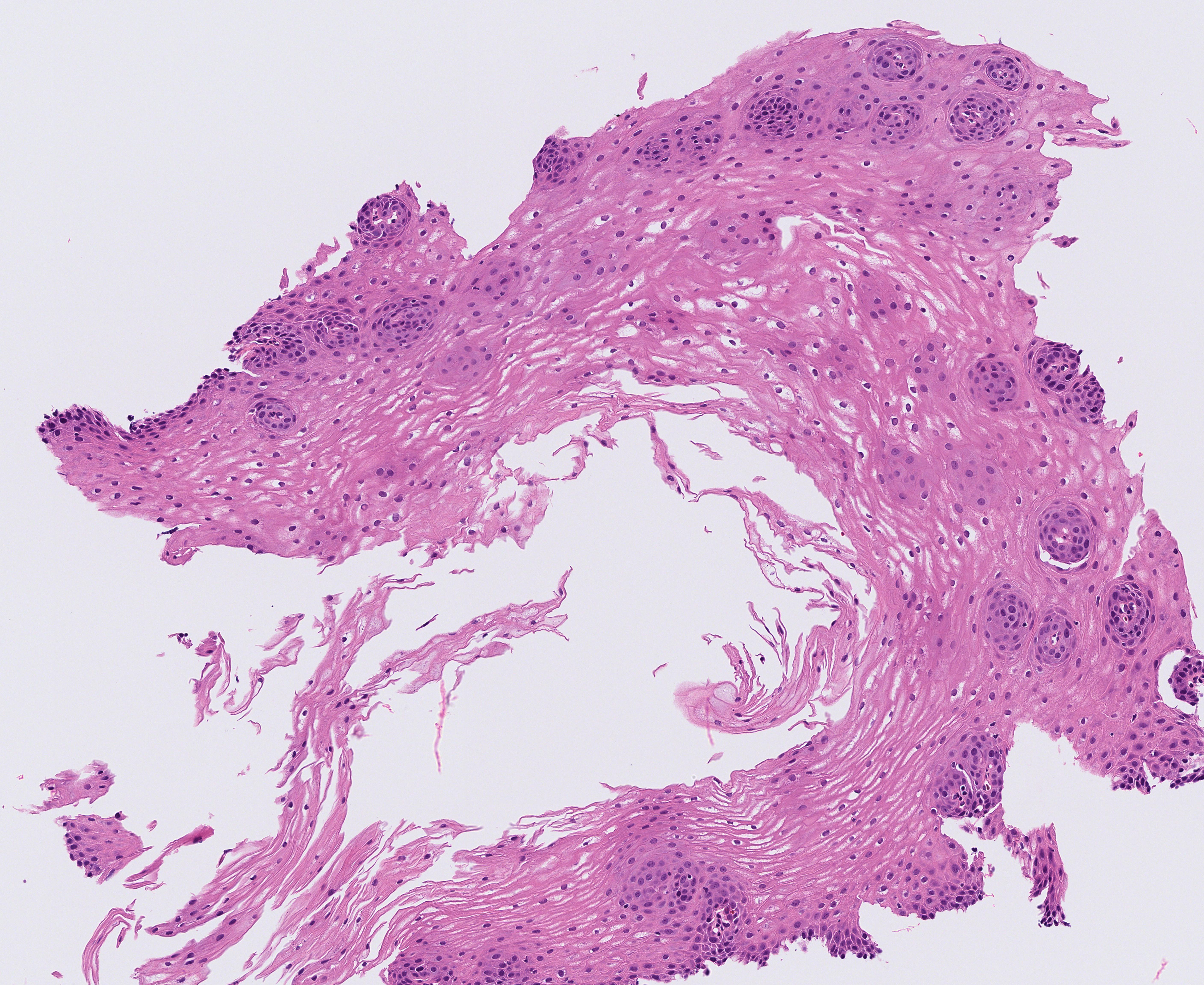}}
  \centerline{(a) Eosinophilic Esophagitis Whole Slide Image}\medskip
\end{minipage}
\label{fig:EoE}
\end{figure}

\section{Related Work}
\label{sec:format}

Given that the size of an WSI can be 100,000x100,000px large, detecting diseases using medical images can be an extremely draining task. Recent medical imaging classification and segmentation efforts have attempted to leverage the strengths of deep learning to assist medical providers, identifying key areas for possible treatment and giving information about a patient's condition. Results have proven to be efficient and effective \cite{Hann2021EnsembleDatasets}\cite{Sarwinda2021DeepCancer}. However, this progress has not properly addressed one of deep learning's most glaring issues: training. Traditional models require large amounts of data to train which may be difficult or impossible in the field of medical image diagnosis. Additionally, the training time needed for traditional models can grow significantly scaling dangerously with the size of the training dataset. Approaches such as domain adaptation have shown promising results by using source and target domains to minimize the training constraints. Continuing this progress, diffusion based methods have show impressive results, outperforming domain adaptation approaches in some studies \cite{Muller-Franzes2022DiffusionImages}. 

\section{Methodology}
\subsection{Preliminary Work}
As with many other medical datasets, the concern for the UVA Medical Center EoE data is that it may not be sufficient for training and any model trained on the datasets can have issues with model generalization. The dataset only contains 30 patients and 514 labeled images/masks, motivating the effort to not use traditional deep learning approaches. Before importing the data into a model and extracting the features, we compared the image distributions of all 30 patients and visualized the results through a network diagram shown in Figure (b). The network is shown with the most unique patients towards the edges of the diagram and the most unique patients in the center. Although all patients were ultimately diagnosed withe EoE at UVA Medical Center, we color coded the patients who's labeled images/masks would meet the "15 or more eosinophil" diagnostic criteria. Additionally, we calculated the aleatoric uncertainty for each patient and represented this through the circle sizes. This was done through a Monte Carlo Dropuout UNet approximation to the deep Gaussian process used in Bayesian Neural Networks \cite{Gal2015BayesianInference}. Minimizing the KL divergence using an approximation through Monte Carlo integration to gives an unbiased estimator. The uncertainty is then given by the following equation \cite{Gal2016DropoutLearning}.
\begin{align*}
    E_{p(z|D)}&H[p(y|z,x)] = \\
    - &\int p(z|D)\left(\sum_{y\in Y} p(y|z,x) \log p(y|z,x)dw\right)
\end{align*}

Immediately apparent was that there were some serious differences in patients. Two quick examples were that patient E-139 had more than 50 times the uncertainty of patient E-28 and that patient E-17 had 71 labelled biopsy images while most patients had only 10 labelled biopsy images. Overall, we can also conclude that the higher the uncertainty, the less unique a patient is relative to the group. These significant differences gives an indication that our dataset is far from homogenous and that traditional deep learning techniques may perform poorly. Thus, domain adaptation and diffusion approaches are a natural progression from the information gathered in our deep dive into the patient data and the recent strong performance compared to domain adaptation techniques. This paper presents progression in both domain adapatation and diffusion based approaches in medical imaging. 
\begin{figure}[htb]

\begin{minipage}[b]{1.0\linewidth}
  \centering
  \centerline{\includegraphics[width=8.5cm]{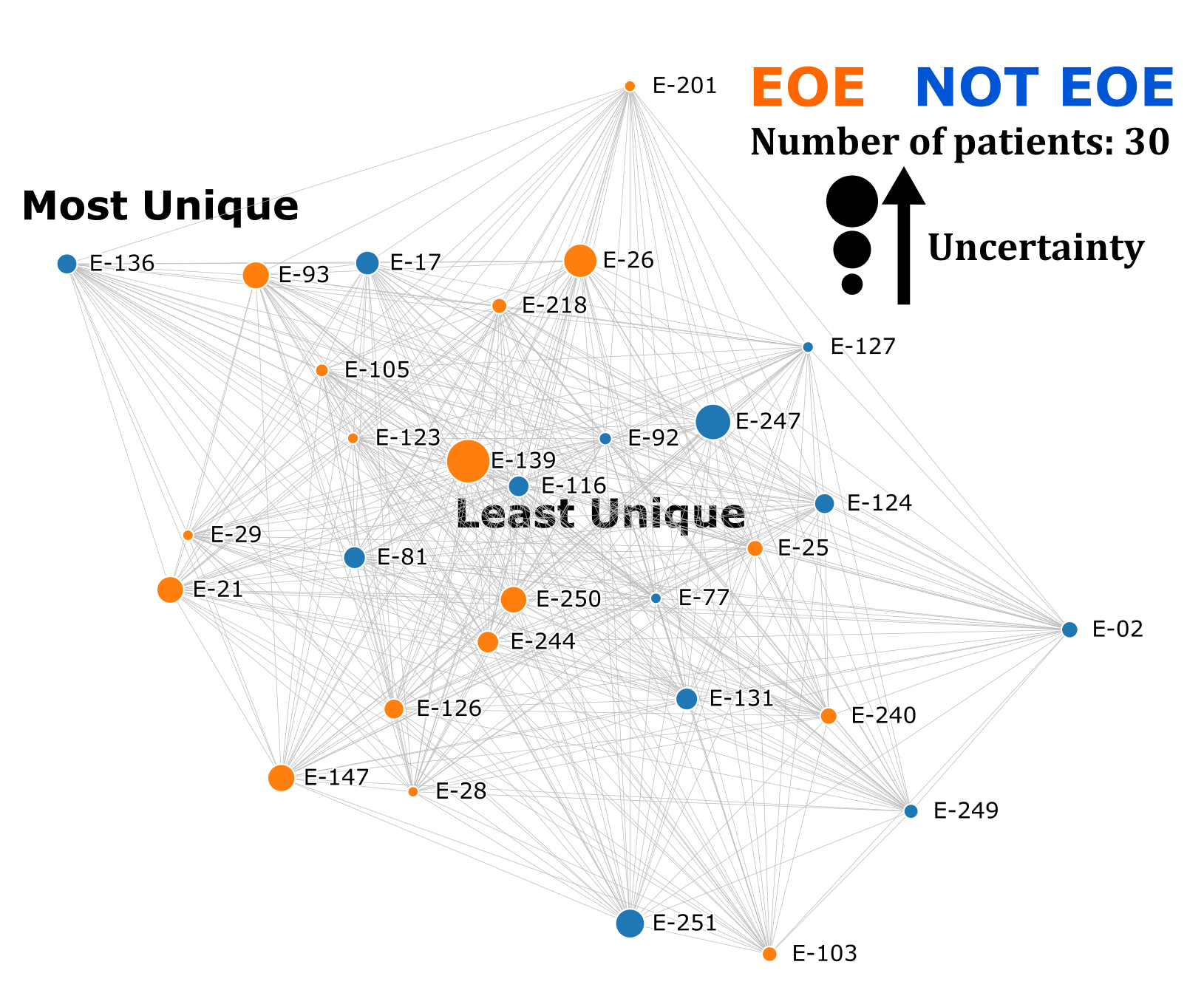}}
  \centerline{(b) EoE Dataset Network Diagram}\medskip
\end{minipage}
\label{fig:net}
\end{figure}

\subsection{Domain Adaptation and Diffusion Based Methods}

For our domain adaptation approach, we split our dataset into three different domains: low, medium, and high uncertainty patients. Since we have 30 patients, each domain will have 10 patients. We will then train each of the combinations and report the average performance. Letting $D_T$ be the target domain and $D_{S_i}$ be the source domain over $X$, then the generalization bound is given by 
\begin{figure}[htb]

\begin{minipage}[b]{1.0\linewidth}
  \centering
  \centerline{\includegraphics[width=8.5cm]{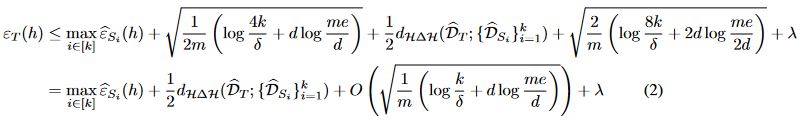}}
  \centerline{(c) MultiDomain Adversarial Network Generalization Bound}\medskip
\end{minipage}
\label{fig:mdanbound}
\end{figure}

\begin{figure}[htb]

\begin{minipage}[b]{1.0\linewidth}
  \centering
  \centerline{\includegraphics[width=8.5cm]{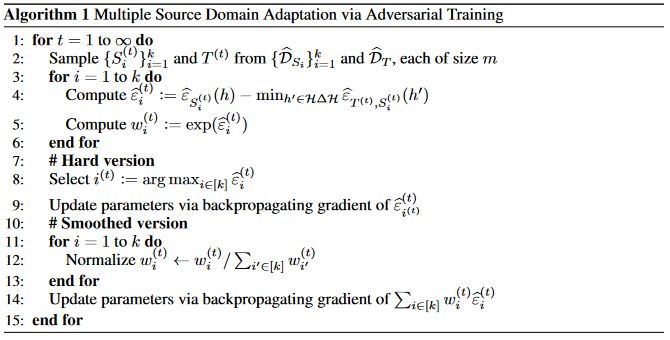}}
  \centerline{(d) MultiDomain Adversarial Network Algorithm}\medskip
\end{minipage}
\label{fig:mdan}
\end{figure}
The goal in the multidomain adversarial network approach is to minimize the generalization bound shown in Figure (c). This is typically done through a minimax saddle point problem and then optimized through adversarial learning

Diffusion based models use variational inference to produce samples matching the data given sufficient time. This is typically done through a parameterized Markov chain that gradually adds noise to the image until the signal is destroyed. A common checkpoint in these approaches is to make sure that the noise indeed reduces the signal to noise ratio to zero (or close to it). In comparison to other deep learning techniques, diffusion models are straightforward to define and efficient to train but there has been limited demonstration that they are capable of generating high quality samples. We will use the following implementation shown in Algorithm~\ref{alg:diff} to apply this diffusion based approach to the EoE dataset\cite{Ho2020DenoisingModels}.  This baseline approach will also be used to compare performance for both the multidomain adversarial network and diffusion approaches. For our baseline, we will use a Monte Carlo Dropout UNet with a dropout value set to 0.5. All training was done on 4 NVIDIA A100 GPUs with 300GB of RAM in TensorFlow/Keras 2.7/PyTorch 2.1.1. Each model run was ran for 100 epochs with a learning rate of 0.001. For the domain adaptation and the diffusion based models, we will produce a Fr\'echet inception distance comparing the generated and real images. For all three models, we will compare the precision and recall. 

\begin{algorithm}
\caption{Training}\label{alg:diff}
\begin{algorithmic}[1]
\Repeat
\\$x_0 \sim q(x_0)$
\\$t \sim$  \text{Uniform}$({1,\dots,T})$
\\$\epsilon \sim N(\textbf{0,I})$
\\Take gradient descent step on
\State $\nabla_{\theta}||\epsilon-\epsilon_{\theta}(\sqrt{\bar{\alpha}}x_0 + \sqrt{1-\bar{\alpha}}\epsilon,t)||^2$
\Until converged
\end{algorithmic}
\end{algorithm}

\section{Results}

\begin{table}[htbp]
  \centering
    \begin{tabular}{|c|c|c|c|c|}
    \hline
    \textbf{Model} & \textit{\textbf{FID}} & \textit{\textbf{Precision}} & \textit{\textbf{Recall}} \\
    \hline
    MCD UNet& N/A & 0.567 & 0.495 \\
    \hline
    MDAN& 42.56 & 0.623 & 0.657\\
    \hline
    DDPM \cite{Ho2020DenoisingModels}& 50.65 & 0.489 & 0.457\\
    \hline
    \end{tabular}%
  \caption{Model Perfomance}
  \label{tab:res}%
\end{table}%

Although there are a few papers \cite{Muller-Franzes2022DiffusionImages} stating that the Diffusion Probabilistic Models beat GANs on medical 2d images, our results suggest that there is at least a space where the GANs may outperform Diffusion methods. In addition, these papers also have cases where some GANs performed better than diffusion methods as shown in Figure (e). We can also see that the precision and recall metrics for the domain adaptation approach are better than both the diffusion and the baseline UNet approach. 

\begin{figure}[htb]
\begin{minipage}[b]{1.0\linewidth}
  \centering
  \centerline{\includegraphics[width=8.5cm]{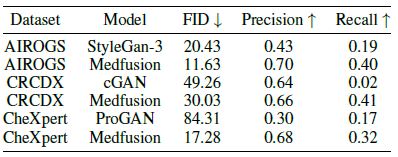}}
  \centerline{(e) Diffusion vs GAN \cite{Muller-Franzes2022DiffusionImages}}\medskip
\end{minipage}
\label{fig:diffgan}
\end{figure}
\begin{figure}[htb]

As a check for the diffusion based methods, we visualized the EoE images after diffusion shown in Figure (f). Clearly, the signal is completely destroyed and there is no indication of cellular structures anywhere in the images. Therefore, the fact that the diffusion based method was able to extract information from this image being in such a state is impressive. 

\begin{minipage}[b]{1.0\linewidth}
  \centering
  \centerline{\includegraphics[width=8.5cm]{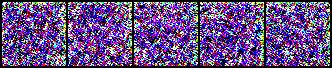}}
  \centerline{(f) Destruction of the EoE Signal}\medskip
\end{minipage}
\label{fig:diffused}
\end{figure}

\section{Conclusion}
The strong performance of the domain adaptation methods compared to the diffusion and the standard UNet approach motivates further work in the domain adaptation field. Next steps could be to see which of the three domains used were most impactful to the relatively strong performance and what impact, if any, would their removal from the dataset be. Additionally, one of the advantages in using a domain adaptation approach is exploring the relationships between different source and target domains. Adding different datasets such as the public PanNuke or Lizard dataset to the analysis can provide different insights into the EoE patients here at UVA and potentially help other researchers in the field of medical image diagnosis.

\begin{acks}
The authors would like to thank the medical researchers of the UVA Gastroenterology (GI) Data Science Laboratory for obtaining the EoE dataset through biopsies of UVA Medical Center patients. Additionally, the staff of the GI Data Science Laboratory tirelessly provided direct medical feedback on our results. Research reported in this publication was supported by National Institutes of Health (NIH) through the National Institute of Diabetes and Digestive and Kidney Diseases (NIDDK) under award numbers K23DK117061-01A1 (Syed) and R01DK132369 (Syed).
\end{acks}

\bibliographystyle{ACM-Reference-Format}
\bibliography{references}

\appendix

\end{document}